\documentclass[11pt]{amsart}

\usepackage[T1]{fontenc}
\usepackage{lmodern}
\usepackage{microtype}
\usepackage[a4paper,margin=1.08in]{geometry}
\usepackage{amsmath,amssymb,amsthm,mathtools}
\usepackage{booktabs}
\usepackage{array}
\usepackage{enumitem}
\usepackage{xcolor}
\usepackage[colorlinks=true,linkcolor=blue!55!black,citecolor=blue!55!black,urlcolor=blue!60!black]{hyperref}

\allowdisplaybreaks

\newtheorem{theorem}{Theorem}[section]
\newtheorem{proposition}[theorem]{Proposition}
\newtheorem{lemma}[theorem]{Lemma}

\theoremstyle{definition}
\newtheorem{definition}[theorem]{Definition}
\newtheorem{problem}[theorem]{Problem}
\newtheorem{remark}[theorem]{Remark}

\newcommand{\cR}{\mathcal R}
\newcommand{\C}{\mathbb C}
\newcommand{\PP}{\mathbb P}
\newcommand{\OO}{\mathcal O}

\newcommand{\im}{\operatorname{im}}
\newcommand{\rk}{\operatorname{rk}}
\newcommand{\codim}{\operatorname{codim}}
\newcommand{\diver}{\operatorname{div}}
\newcommand{\Hom}{\operatorname{Hom}}

\newcommand{\bfA}{\mathbf A}
\newcommand{\bfB}{\mathbf B}
\newcommand{\calA}{\mathcal A}
\newcommand{\calB}{\mathcal B}

\title{Integral zeros of Krawtchouk polynomials and quantum entanglements}

\author{Jaeseong Heo}
\address{Department of Mathematics, Research Institute for Natural Sciences, Hanyang University, Seoul 04763, Korea}
\email{hjs@hanyang.ac.kr}

\author{Young-Hoon Kiem}
\address{School of Mathematics, Korea Institute for Advanced Study, 85 Hoegiro, Dongdaemun-gu, Seoul, Korea}
\email{kiem@kias.re.kr}

\author{Chul-hee Lee$^*$}
\address{Center for AI and Natural Sciences, Korea Institute for Advanced Study, 85 Hoegiro, Dongdaemun-gu, Seoul, Korea}
\email{chlee@kias.re.kr}
\thanks{$^*$ Corresponding author}

\date{}

\begin{document}

\begin{abstract}
In \cite{HeoKiem2019}, the authors raised a problem relating integral zeros of binary Krawtchouk polynomials, parametrized by quadruples $(m,n,k,\ell)$ with $m\le n$ and $k\le \ell$, to the existence of pairs of subspaces satisfying the range criterion in quantum information theory, and settled the symmetric case $k=\ell$. In this paper, we solve the problem for the following three asymmetric families:
\begin{align*}
 (m,n,k,\ell)&=\Bigl(3,\ t^2-1,\ \tfrac{t(t-1)}2,\ \tfrac{t(t+1)}2\Bigr), && t\ge2,\\
 (m,n,k,\ell)&=\Bigl(4,\ 3t^2+2t-1,\ \tfrac{3t^2-t}2,\ \tfrac{3t^2+5t+2}2\Bigr), && t\ge2,\\
 (m,n,k,\ell)&=\Bigl(4,\ 3s^2+4s,\ \tfrac{3s^2+s}2,\ \tfrac{3s^2+7s+4}2\Bigr), && s\ge1.
\end{align*}
The remaining cases are sporadic zeros, with only finitely many for each $m\ge 5$.
\end{abstract}

\maketitle


\section{Introduction}

In quantum information, a key challenge is determining whether a quantum state is separable or entangled. The PPT (positive partial transpose) criterion \cite{Peres1996} is a necessary condition for separability, which says that if a state is separable, its partial transpose is also positive semidefinite. The range criterion for separability \cite{Horodecki1997} tells us that if a (mixed) quantum state $A$ acting on $\C^m\otimes \C^n$ is separable,
there exists a family $\{x_i \otimes y_i\}$ of product vectors such that
$$
\cR(A) = \text{span}\{x_i \otimes y_i\} ~~\text{and}~~ \cR(A^\tau)=\text{span}\{\bar x_i\otimes y_i\}
$$
where $A^\tau$ denotes the partial transpose and $\bar{x}$ is obtained by taking the complex conjugates of the coordinates of $x$. 
In particular, for a separable state $A$, there should be a nonzero vector 
\begin{equation}\label{RangeMot} 
x\otimes y\in \cR(A) \quad \text{with}\quad \bar x\otimes y\in \cR(A^\tau).\end{equation}

A \emph{PPTES} is an entangled state that satisfies the PPT criterion. 
An \emph{edge state} $A$ is a PPTES that fails the range criterion even in the weaker form of \eqref{RangeMot}, i.e. there is no non-zero product vector satisfying \eqref{RangeMot}.
It is well-known \cite{LKHC2001} that every PPTES can be expressed as a convex sum of separable states and edge states and hence classifying edge states is essential to understanding all PPTES, as they form key building blocks. 

The first step towards finding all edge states is to find subspaces $D, E\subset \C^m\otimes\C^n$ of codimensions $k$ and $\ell$ respectively, 
which may be the ranges 
$$(D,E) = (\cR(A),\cR(A^\tau))$$
for an edge state $A$.
Then the weaker form \eqref{RangeMot} of the range criterion raises the following.
\begin{problem}[Range-criterion problem \cite{HeoKiem2019,KiemKyeLee2011}]\label{RangeProb} 
   Find all quadruples $(m,n,k,\ell)$ of natural numbers for which there exist subspaces
\[
 D,E\subset \C^m\otimes\C^n,\quad \codim D=k,\quad \codim E=\ell,
\]
such that
\begin{equation}\label{eq:range-condition}
x\otimes y\in D,\quad \overline{x}\otimes y\in E
\quad\Longrightarrow\quad
x=0\ \text{or}\ y=0
\qquad
(x\in\C^m,\ y\in\C^n).
\end{equation} 
\end{problem}
By symmetry, we may and will assume $m\le n$ and $k\le \ell$ from now on.

By \cite{KiemKyeLee2011}, a key condition in this problem is expressed in terms of binary Krawtchouk polynomials.
For integers \(N\ge0\) and \(0\le q\le N\), the binary Krawtchouk polynomial \(P_q^N(x)\) is defined by 
\[ P_q^N(x) = \sum_{j=0}^q (-1)^j \binom{x}{j} \binom{N-x}{q-j}. \]

The range-criterion problem can then be divided into the following four cases:
\begin{enumerate}[label=\textup{(\alph*)}]
\item\label{case:a} \(k+\ell<m+n-2\);
\item\label{case:b} \(k+\ell>m+n-2\);
\item\label{case:c} \(k+\ell=m+n-2\) and \(P_{m-1}^{k+\ell}(k)\ne0\);
\item\label{case:d} \(k+\ell=m+n-2\) and \(P_{m-1}^{k+\ell}(k)=0\).
\end{enumerate}
Kiem, Kye and Lee \cite{KiemKyeLee2011} proved that such subspaces do not exist in case~\ref{case:a}, whereas they exist in case~\ref{case:b}.
They also showed non-existence in case~\ref{case:c} via a cohomological obstruction.
The remaining case is~\ref{case:d}, which may be regarded as the critical regime.

Heo and Kiem \cite{HeoKiem2019} conjectured that such subspaces \(D\) and \(E\) exist for every quadruple $(m,n,k,\ell)$ in case~\ref{case:d}, and proved the conjecture in the following cases \cite[Theorem~1.3]{HeoKiem2019}:
\begin{itemize} \setlength{\itemsep}{-.5pt}
\item[(i)] $m=2$ or $n=2$,  
\item[(ii)] $k=1$ or $\ell=1$, 
\item[(iii)] $k=\ell$.
\end{itemize}
We call the case \(k=\ell\) symmetric and the case \(k\ne\ell\) asymmetric.
For \(m=3\) and \(m=4\), the asymmetric positive integral solutions in case~\ref{case:d} form the following three families:
\begin{equation}
m=3,\quad
n=t^2-1,\quad
k=\frac{t(t-1)}{2},\quad
\ell=\frac{t(t+1)}{2},
\qquad t\ge2,
\label{eq:m3-family}
\end{equation}
\begin{equation}
m=4,\quad
n=3t^2+2t-1,\quad
k=\frac{3t^2-t}{2},\quad
\ell=\frac{3t^2+5t+2}{2},
\qquad t\ge2,
\label{eq:m4-family}
\end{equation}
and
\begin{equation}
m=4,\quad
n=3s^2+4s,\quad
k=\frac{3s^2+s}{2},\quad
\ell=\frac{3s^2+7s+4}{2},
\qquad s\ge1.
\label{eq:m4-second-family}
\end{equation}
Together with the symmetric \(m=4\) family, these exhaust the positive integral quadruples in case~\ref{case:d} for \(m=3\) and \(m=4\); see Proposition~\ref{prop:low-degree-krawtchouk}.
For \(m\ge5\), by contrast, only finitely many integers \(n\) can occur in case~\ref{case:d} for some nontrivial \(k,\ell\); see \cite{Kra92} and \cite[Theorem~2]{Kra96}.

\medskip

Our main theorem is as follows.
\begin{theorem}\label{thm:main}
For every quadruple \((m,n,k,\ell)\) belonging to one of the three families \eqref{eq:m3-family}, \eqref{eq:m4-family}, and \eqref{eq:m4-second-family}, there exist subspaces \(D,E\subset \C^m\otimes\C^n\) of codimensions \(k\) and \(\ell\), respectively, satisfying the range condition \eqref{eq:range-condition}.
\end{theorem}

The proof naturally divides into three parts, reflected in the organization of the paper.
In Section~\ref{sec:background}, we reformulate the range condition in terms of matrices and establish the preliminary results concerning the relevant Krawtchouk zeros.
In Section~\ref{sec:m3}, we prove Theorem~\ref{thm:main} for the family \eqref{eq:m3-family} by an elementary and explicit construction, realizing the \(n=t^2-1\) dimensional second factor as the kernel of the divergence map on triples of homogeneous polynomials in three variables.
In Section~\ref{sec:m4}, we treat the two \(m=4\) families \eqref{eq:m4-family} and \eqref{eq:m4-second-family} with \(t\ge3\) and \(s\ge2\), respectively.
We first establish a vector-bundle criterion showing that a resolution
\[
0\longrightarrow \OO_{\PP^3}(-2)^k
\longrightarrow \OO_{\PP^3}(-1)^n
\longrightarrow \OO_{\PP^3}^{\ell}
\longrightarrow F\longrightarrow 0
\]
of a vector bundle $F$ on the projective space $\PP^3$ produces the required pair \((D,E)\).
We then construct such resolutions using the existence of stable rank-two bundles with natural cohomology from \cite{HartshorneHirschowitz1982}.
The two \(m=4\) families correspond to the two residue classes arising from the single uniform construction of Theorem~\ref{thm:m4-delta}.
The two smallest members, namely \(t=2\) in \eqref{eq:m4-family} and \(s=1\) in \eqref{eq:m4-second-family}, fall outside this uniform construction.
In Section~\ref{sec:4727-witness}, we treat these two cases separately by explicit matrix constructions.

\section{Background}\label{sec:background}
\subsection{Matrix formulation of the range condition}\label{subsec:linear-algebra}
Let \(D\subset \C^m\otimes\C^n\) and \(E\subset \C^m\otimes\C^n\) have codimensions \(k\) and \(\ell\), respectively.  Write \(D=\ker\calA\) and \(E=\ker\calB\), where \(\calA:\C^m\otimes\C^n\to\C^k\) and
\(\calB:\C^m\otimes\C^n\to\C^\ell\) are surjective linear maps.

Let \(e_1,\dots,e_m\) be the standard basis of \(\C^m\).  
Define \(A_i\in M_{k\times n}(\C)\) and \(B_i\in M_{n\times\ell}(\C)\) by \(\calA(e_i\otimes y)=A_i y\) and
\(\calB(e_i\otimes y)=B_i^*y\).  
For \(x=\sum_i x_i e_i\), put \(A(x):=\sum_i x_iA_i\) and \(B(x):=\sum_i x_iB_i\).  
Then \(\calA(x\otimes y)=A(x)y\) and \(\calB(\overline{x}\otimes y)=B(x)^*y\), and hence
\begin{equation}\label{eq:membership-conditions}
x\otimes y\in D \iff A(x)y=0,
\qquad
\overline{x}\otimes y\in E \iff B(x)^*y=0.
\end{equation}
Set
\begin{equation}\label{eq:block-matrices}
\bfA=\begin{bmatrix}A_1&\cdots&A_m\end{bmatrix}
\in M_{k\times mn}(\C),
\qquad
\bfB=\begin{bmatrix}
B_1\\
\vdots\\
B_m
\end{bmatrix}
\in M_{mn\times\ell}(\C).
\end{equation}
Then
\[
\im \bfA
=
\sum_{i=1}^m \im A_i,
\qquad
\ker\bfB
=
\bigcap_{i=1}^m\ker B_i.
\]
Hence \(\calA\) and \(\calB\) are surjective precisely when
\begin{equation}\label{eq:surjectivity-conditions}
\im\bfA=\C^k,
\qquad
\ker\bfB=0,
\end{equation}
or equivalently when
\begin{equation}\label{eq:block-rank-conditions}
\rk\bfA=k,
\qquad
\rk\bfB=\ell.
\end{equation}

Conversely, any matrices \(A_i\in M_{k\times n}(\C)\) and \(B_i\in M_{n\times\ell}(\C)\) satisfying the equivalent conditions \eqref{eq:surjectivity-conditions} and \eqref{eq:block-rank-conditions} define subspaces \(D\) and \(E\) of codimensions \(k\) and \(\ell\), respectively, through the formulas above.

We can formulate the range condition entirely in terms of \(A(x)\) and \(B(x)\).

\begin{proposition}
\label{prop:kernel-intersection}
With the notation above, the range condition \eqref{eq:range-condition} is equivalent to
\[
\ker A(x)\cap\ker B(x)^*=0
\qquad
\text{for every }x\in\C^m\setminus\{0\}.
\]
\end{proposition}

\begin{proof}
By \eqref{eq:membership-conditions}, for \(x\ne0\) there exists a nonzero \(y\) violating \eqref{eq:range-condition} if and only if
\[
\ker A(x)\cap\ker B(x)^*\ne0.
\]
Hence the result follows.
\end{proof}

\subsection{Integral zeros of Krawtchouk polynomials for \(m=3\) and \(m=4\)}
\begin{proposition}\label{prop:low-degree-krawtchouk}
Let $m\in\{3,4\}$ and let $n,k,\ell$ be positive integers.
Then
\[
k+\ell=m+n-2,
\qquad
P_{m-1}^{k+\ell}(k)=0
\]
if and only if, up to interchanging $k$ and $\ell$, one of the following holds:
\begin{enumerate}
\item[\textup{(i)}] \((m,n,k,\ell)=\left(3,\,t^2-1,\,\frac{t(t-1)}2,\,\frac{t(t+1)}2\right)\) for some \(t\ge2\).
\item[\textup{(ii)}] \((m,n,k,\ell)=(4,\,2t-2,\,t,\,t)\) for some \(t\ge2\).
\item[\textup{(iii)}] \((m,n,k,\ell)=\left(4,\,3t^2+2t-1,\,\frac{3t^2-t}{2},\,\frac{3t^2+5t+2}{2}\right)\) for some \(t\ge1\).
\item[\textup{(iv)}] \((m,n,k,\ell)=\left(4,\,3s^2+4s,\,\frac{3s^2+s}{2},\,\frac{3s^2+7s+4}{2}\right)\) for some \(s\ge1\).
\end{enumerate}
\end{proposition}

\begin{proof}
Assume that $m=3$ and that $n,k,\ell$ are positive integers satisfying
\(k+\ell=m+n-2\) and \(P_{m-1}^{k+\ell}(k)=0\).
Then \(n= k+\ell-1\) and
\[
P_2^{k+\ell}(k)
=
\binom{\ell}{2}-k\ell+\binom{k}{2}
=
\frac{(\ell-k)^2-(k+\ell)}{2}.
\]
Hence \(P_2^{k+\ell}(k)=0\) if and only if
\((\ell-k)^2=k+\ell=n+1\).

Since the equation is symmetric in $k$ and $\ell$, we may assume
\(k\le\ell\). The case \(k=\ell\) is impossible, so put
\(t=\ell-k>0\). Then \(k+\ell=t^2\), and consequently
\[
k=\frac{t(t-1)}{2},
\qquad
\ell=\frac{t(t+1)}{2},
\qquad
n=t^2-1.
\]
Since \(k>0\), we have \(t\ge2\). Conversely, these values satisfy
both \(k+\ell=m+n-2\) and \(P_2^{k+\ell}(k)=0\).

Assume that $m=4$ and that $n,k,\ell$ are positive integers satisfying
\(k+\ell=m+n-2\) and \(P_{m-1}^{k+\ell}(k)=0\).
Then \(n=k+\ell-2\) and
\[
P_3^{k+\ell}(k)
=
\frac{(\ell-k)\bigl((\ell-k)^2-3(k+\ell)+2\bigr)}{6}.
\]
Hence \(P_3^{k+\ell}(k)=0\) if and only if either \(k=\ell\), or
\[
(\ell-k)^2=3(k+\ell)-2=3n+4.
\]

The case \(k=\ell\) gives the symmetric family.
We therefore consider the asymmetric case. 
Since the equation is symmetric in $k$ and $\ell$, we may assume \(k<\ell\)
and put \(\delta=\ell-k>0\). Then
\(\delta^2=3(k+\ell)-2\), so
\[
k+\ell=\frac{\delta^2+2}{3},
\qquad
n=\frac{\delta^2-4}{3}.
\]
In particular, \(3\nmid\delta\). Solving for $k$ and $\ell$ gives
\[
k=\frac{(\delta-1)(\delta-2)}{6},
\qquad
\ell=\frac{(\delta+1)(\delta+2)}{6}.
\]
Since \(k>0\), we have \(\delta\ge4\).

There are two possibilities modulo $3$. If \(\delta=3t+1\), then
\[
n=3t^2+2t-1,
\qquad
k=\frac{3t^2-t}{2},
\qquad
\ell=\frac{3t^2+5t+2}{2},
\qquad t\ge1.
\]
If \(\delta=3s+2\), then
\[
n=3s^2+4s,
\qquad
k=\frac{3s^2+s}{2},
\qquad
\ell=\frac{3s^2+7s+4}{2},
\qquad s\ge1.
\]
Conversely, both families satisfy \(k+\ell=m+n-2\) and
\(P_3^{k+\ell}(k)=0\).
\end{proof}

Among the four families in Proposition~\ref{prop:low-degree-krawtchouk}, case {\rm (ii)} is the symmetric case \(k=\ell\), and is already covered by \cite[Theorem~1.3]{HeoKiem2019} as discussed in the Introduction.
In case {\rm (i)}, the value \(t=2\) gives \(k=1\), and in case {\rm (iii)}, the value \(t=1\) also gives
\(k=1\); these are again covered by the same theorem.

Thus, the cases not covered by the previous results are case {\rm (i)} with \(t\ge3\), case {\rm (iii)} with \(t\ge2\), and case {\rm (iv)} with \(s\ge1\).

\section{An explicit construction for the $m=3$ family}\label{sec:m3}
We first treat the \(m=3\) family \eqref{eq:m3-family}.
Throughout this section, let \(t\ge2\), set \(r=t-2\), and write \(n=t^2-1\), \(k=t(t-1)/2\), and \(\ell=t(t+1)/2\).
In this case, the required subspaces \(D\) and \(E\) can be constructed explicitly using divergence-free homogeneous polynomials in three variables.
Write \(S_d=\C[z_1,z_2,z_3]_d\) for the homogeneous polynomials of degree \(d\), with the convention \(S_d=0\) for \(d<0\).
Write
\[
 \partial_i=\frac{\partial}{\partial z_i},
 \qquad
 z=(z_1,z_2,z_3).
\]
Define
\[
 W_r=
 \left\{
 f=(f_1,f_2,f_3)\in S_r^3:
 \diver f=\partial_1f_1+\partial_2f_2+\partial_3f_3=0
 \right\}.
\]

\begin{lemma}\label{lem:Wr-dim}
The divergence map \(\diver:S_r^3\longrightarrow S_{r-1}\) is surjective,
and \(\dim W_r=t^2-1\).
\end{lemma}

\begin{proof}
If \(r=0\), then \(S_{r-1}=0\), so the divergence map is trivially surjective.
Assume now that \(r\ge1\).
For $q\in S_{r-1}$, consider \(zq=(z_1q,z_2q,z_3q)\in S_r^3\).
Since $q$ is homogeneous of degree $r-1$, Euler's identity gives
\[
 \sum_{i=1}^3 z_i(\partial_iq)=(r-1)q.
\]
Therefore
\[
 \diver(zq)
 =\sum_{i=1}^3\partial_i(z_iq)
 =3q+\sum_{i=1}^3 z_i(\partial_iq)
 =(r+2)q.
\]
Since $r+2=t \ne 0$, the divergence map is surjective.
Consequently,
\[
 \dim W_r=\dim S_r^3-\dim S_{r-1} = 3\binom{r+2}{2}-\binom{r+1}{2}=t^2-1.
\]
\end{proof}

By Lemma~\ref{lem:Wr-dim},
\[
\dim W_r=n,\qquad
\dim S_r=k,\qquad
\dim S_{r+1}=\ell.
\]
For \(1\le i\le3\), define linear maps \(A_i:W_r\to S_r\) and
\(B_i^*:W_r\to S_{r+1}\) by
\[
A_i f=f_i,
\qquad
B_i^* f
=
\det\!\begin{pmatrix}
z_1&z_2&z_3\\
\delta_{i1}&\delta_{i2}&\delta_{i3}\\
f_1&f_2&f_3
\end{pmatrix},
\]
where \(\delta_{ij}\) denotes the Kronecker delta.
Then, with the notation of Subsection~\ref{subsec:linear-algebra},
\begin{align*}
A(x)f
&=x_1f_1+x_2f_2+x_3f_3,
\\
B(x)^*f
&=
\det\!\begin{pmatrix}
z_1&z_2&z_3\\
\overline{x}_1&\overline{x}_2&\overline{x}_3\\
f_1&f_2&f_3
\end{pmatrix}.
\end{align*}
These define linear maps
\[
\calA:\C^3\otimes W_r\to S_r,
\qquad
\calB:\C^3\otimes W_r\to S_{r+1},
\]
by \(\calA(x\otimes f)=A(x)f\) and \(\calB(x\otimes f)=\sum_{i=1}^3x_iB_i^*f\).
In particular,
\[
\calB(\overline{x}\otimes f)
=
\sum_{i=1}^3\overline{x_i}B_i^*f
=
B(x)^*f.
\]
We then set
\begin{equation}\label{eq:DE-m3}
D=\ker\calA,
\qquad
E=\ker\calB.
\end{equation}

\begin{lemma}\label{lem:AB-surj}
The maps $\calA$ and $\calB$ are surjective.
Consequently,
\[
\codim D=\dim S_r=k,
\qquad
\codim E=\dim S_{r+1}=\ell.
\]
\end{lemma}

\begin{proof}
Let \(P\in S_r\).
Choose \(f_2\in S_r\) such that \(\partial_2f_2=-\partial_1P\), and set \(f=(P,f_2,0)\).
Then \(f\in W_r\) and \(\calA(e_1\otimes f)=A(e_1)f=P\).
Hence \(\calA\) is surjective.

To prove that \(\calB\) is surjective, it is enough to obtain every
monomial \(M\in S_{r+1}\).
Choose \(c\) such that \(z_c\mid M\), and choose distinct \(a,b\) with \(\{a,b,c\}=\{1,2,3\}\).
Set \(f_c=0\) and choose \(f_b=\pm M/z_c\), with the sign chosen so that \(B_a^*f=M\).
Choose \(f_a\in S_r\) such that \(\partial_af_a=-\partial_bf_b\).
Then \(f\in W_r\) and \(\calB(e_a\otimes f)=B_a^*f=M\).
Since the monomials span \(S_{r+1}\), \(\calB\) is surjective.
\end{proof}

\begin{lemma}\label{lem:m3-kernel}
Let $x\in\C^3\setminus\{0\}$ and $f\in W_r$.  
If \(A(x)f=0\) and \(B(x)^*f=0\), then \(f=0\).
\end{lemma}

\begin{proof}
To simplify the two vanishing conditions simultaneously, we choose a basis adapted to the roles of \(x\) and \(\overline{x}\).
Let $V=\C^3$, set \[ \xi:=x^{\mathsf T}\in V^*, \qquad c:=\xi(\overline{x})=\sum_{i=1}^3|x_i|^2>0.\]
Thus, $\overline{x}\notin\ker\xi$.
Choose a basis $v_2,v_3$ of $\ker\xi$ and set \(v_1=\overline{x}\).
Then $v_1,v_2,v_3$ is a basis for $V$.
Let $w_1,w_2,w_3$ denote the corresponding dual basis.
Since $v_2,v_3\in\ker\xi$ and $\xi(v_1)=c$, we have \(\xi=cw_1\).

Let $P$ be the change-of-basis matrix whose columns are $v_1,v_2,v_3$.
Then
\[
\begin{pmatrix}
z_1\\
z_2\\
z_3
\end{pmatrix}
=P
\begin{pmatrix}
w_1\\
w_2\\
w_3
\end{pmatrix}.
\]
Define \(\widetilde f(w):=P^{-1}f(Pw)\), so that \(f(Pw)=P\widetilde f(w)\).

By the chain rule, the Jacobian becomes
\[
J_{\widetilde f}(w)
=
P^{-1}J_f(Pw)P.
\]
Taking traces gives
\[
\diver_w\widetilde f(w)
=
(\diver_z f)(Pw).
\]

Since $\xi=cw_1$,
\[
A(x)f(Pw)
=
\xi\bigl(P\widetilde f(w)\bigr)
=
c\,\widetilde f_1(w).
\]

Moreover, since $\overline{x}=v_1=Pe_1$,
\begin{align*}
(B(x)^*f)(Pw)
&=
\det\bigl(Pw,Pe_1,P\widetilde f(w)\bigr)\\
&=
\det(P)\,
\det\bigl(w,e_1,\widetilde f(w)\bigr)\\
&=
\det(P)(w_3\widetilde f_2(w)-w_2\widetilde f_3(w))
.
\end{align*}

Thus, $\diver_z f=0$, $A(x)f=0$ and $B(x)^*f=0$ imply
\[
\diver_w \widetilde f=0,
\qquad
\widetilde f_1=0,
\qquad
w_2\widetilde f_3-w_3 \widetilde f_2=0.
\]

Since $w_2$ and $w_3$ are relatively prime, there is $q\in S_{r-1}=\C[w_1,w_2,w_3]_{r-1}$ such that
\[
\widetilde f_2=w_2q,
\qquad
\widetilde f_3=w_3q.
\]
Using \(\diver_w\widetilde f=0\), we obtain
\[
\diver_w\widetilde f
=\partial_{w_2}(w_2q)+\partial_{w_3}(w_3q)
=\left(2+w_2\partial_{w_2}+w_3\partial_{w_3}\right)q=0.
\]
On a monomial $w_1^aw_2^bw_3^d$, the operator \(2+w_2\partial_{w_2}+w_3\partial_{w_3}\) acts by multiplication by the positive integer
\[
2+b+d.
\]
It is therefore invertible on $\C[w_1,w_2,w_3]_{r-1}$, so $q=0$.  
Hence $\widetilde f=0$, and therefore $f=0$.
\end{proof}

\begin{proof}[Proof of Theorem~\ref{thm:main} for the family \eqref{eq:m3-family}]
By Lemmas~\ref{lem:Wr-dim} and \ref{lem:AB-surj}, the spaces $D$ and $E$ defined in \eqref{eq:DE-m3} have the required codimensions.
If
\(x\otimes f\in D\) and \(\overline{x}\otimes f\in E\) with $x\ne 0$, then Lemma~\ref{lem:m3-kernel} gives $f=0$.
Hence \(D\) and \(E\) satisfy the range condition \eqref{eq:range-condition}.
\end{proof}

\section{A vector-bundle construction for the two $m=4$ families}\label{sec:m4}
In this section, we prove Theorem \ref{thm:main} for the families \eqref{eq:m4-family} and \eqref{eq:m4-second-family} by using a vector bundle technique in algebraic geometry. In Subsection~\ref{sec:criterion}, we reduce the range-criterion problem to the existence of a vector bundle $F$ over the projective space $\PP^3$ and a suitable resolution. In Subsection~\ref{subsec:finite-resolution}, we find a condition for the existence of $F$ which admits a desired resolution. In Subsection~\ref{subsec:natural-cohomology}, we show that such a vector bundle really exists by a classical result of Hartshorne and Hirschowitz \cite{HartshorneHirschowitz1982}.
The \(m=3\) family treated in the previous section also admits a vector-bundle interpretation on \(\PP^2\), where the relevant bundle has rank one and can be described explicitly, whereas the \(m=4\) case involves nontrivial rank-two bundles on \(\PP^3\).

\subsection{Reduction to vector bundles}\label{sec:criterion}
Proposition~\ref{prop:kernel-intersection} reduces the range-criterion problem to producing matrices
\(A_i\in M_{k\times n}(\C)\) and \(B_i\in M_{n\times\ell}(\C)\) satisfying the corresponding linear-algebraic conditions.
This section manufactures such matrices from a different source: a resolution of a vector bundle on projective space.
Since this is first introduced at this point in the paper, we will first establish the correspondence between the two figures before proceeding further.
For precise definitions and basic facts about algebraic geometry, we refer to \cite{Hart, Vakil}. 

Write $X=\PP(V)=\PP^{m-1}_\C$ for the space of lines through the origin in $V=\C^m$: a point of $X$ is a nonzero $x\in V$ up to a nonzero scalar, and the homogeneous coordinates $x_1,\dots,x_m$ are simply the coordinates of $x$ itself.  For each integer $d$, the line bundle $\OO_X(d)$ is, concretely, the bookkeeping device whose global sections $H^0(\OO_X(d))$, for $d\ge0$, are exactly the degree-$d$ homogeneous polynomials in $x_1,\dots,x_m$; in particular, $H^0(\OO_X(1))=V^*$ is spanned by the coordinates $x_1,\dots,x_m$ themselves.  A \emph{vector bundle} $F$ on $X$ is, for our purposes, nothing more than a family of vector spaces $F_x$ varying algebraically with $x\in X$, and \emph{twisting} $F$ by $\OO_X(d)$ (written $F(d):=F\otimes\OO_X(d)$) shifts by $d$ the degree of homogeneity with which $F$ is allowed to depend on the overall scale of $x$.  The one fact this section leans on repeatedly is this: a map of vector bundles between twists one level apart -- such as $\OO_X(-2)^k\to\OO_X(-1)^n$, or $\OO_X(-1)^n\to\OO_X^{\ell}$ -- is exactly a matrix whose entries are \emph{linear forms} in $x_1,\dots,x_m$, no more and no less, and this is precisely the pencil notation $A(x)=\sum_ix_iA_i$ introduced in Subsection~\ref{subsec:linear-algebra}.
Dualizing a bundle reverses arrows and negates twists, $\OO_X(d)^\vee=\OO_X(-d)$, and $H^0(F^\vee)=\Hom(F,\OO_X)$ is the space of bundle maps from $F$ to the untwisted trivial line bundle -- concretely, the space of linear functionals on $F$ that are defined globally and consistently across all of $X$.

With this dictionary in hand, the following proposition converts a suitable linear resolution on $X$ into a solution of the range-criterion problem.

\begin{proposition}\label{prop:resolution-criterion}
Let $V=\C^m$, $X=\PP(V)$, and let $F$ be a vector bundle on $X$ with $H^0(F^\vee)=0$. Suppose that there is an exact sequence
\begin{equation}\label{eq:pure-resolution}
0\longrightarrow \OO_X(-2)^k
\longrightarrow \OO_X(-1)^n
\longrightarrow \OO_X^{\ell}
\longrightarrow F\longrightarrow 0.
\end{equation}
Then there are subspaces \(D,E\subset V\otimes\C^n\) of codimensions \(k\) and \(\ell\), respectively, satisfying \eqref{eq:range-condition}.
\end{proposition}

\begin{proof}
Let $K$ be the image of $\OO_X(-1)^n\to\OO_X^{\ell}$, so that \eqref{eq:pure-resolution} splits into two short exact sequences
\[
 0\to K\to\OO_X^{\ell}\to F\to0,
 \qquad
 0\to\OO_X(-2)^k\to\OO_X(-1)^n\to K\to0 .
\]
Since $F$ is locally free, the first sequence is locally split (near any point it looks, after choosing local trivializations, like a split short exact sequence of vector spaces), so $K$ is again locally free; hence the second sequence is locally split as well.  A locally split short exact sequence of vector bundles stays exact after applying $\Hom(-,\OO_X)$ (dualizing) or after restriction to any single fiber, precisely because a split short exact sequence of vector spaces does both.  Dualizing reverses arrows and negates twists ($\OO_X(d)^\vee=\OO_X(-d)$), so dualizing \eqref{eq:pure-resolution} and then twisting by $\OO_X(-1)$ (which shifts every twist down by one) turns it into an exact sequence in the same range of twists with which this paper works:
\begin{equation}\label{eq:dual-resolution}
0\longrightarrow F^\vee(-1)
\longrightarrow \OO_X(-1)^\ell
\xrightarrow{\,B\,}\OO_X^n
\xrightarrow{\,A\,}\OO_X(1)^k
\longrightarrow 0.
\end{equation}
By the dictionary above, the two bundle maps $A$ and $B$ here -- each between twists one level apart -- are exactly matrices of linear forms: choosing homogeneous coordinates $x_1,\dots,x_m$, write
\[
 A(x)=\sum_{i=1}^m x_iA_i,
 \qquad
 B(x)=\sum_{i=1}^m x_iB_i,
\]
where $A_i\in M_{k\times n}(\C)$ and $B_i\in M_{n\times\ell}(\C)$.

Since \eqref{eq:dual-resolution} is exact, $A$ is surjective on every fiber, so for any nonzero $x$ the matrix $A(x)$ maps onto $\C^k$.  Hence
\(\sum_{i=1}^m\im A_i=\C^k\).

For $B$, the left exactness of global sections for \eqref{eq:dual-resolution} twisted by $\OO_X(1)$ identifies
\[
 H^0(F^\vee)\simeq\ker\!\left(\C^\ell\xrightarrow{\ c\mapsto\sum_i x_iB_ic\ }
 H^0(\OO_X(1))\otimes\C^n\right)
 =\bigcap_i\ker B_i,
\]
since, as noted above, the homogeneous coordinates form a basis of $H^0(\OO_X(1))$.  
As $H^0(F^\vee)=0$ by assumption, we have 
\(\bigcap_i\ker B_i=0\).

Now fix $x\ne 0$ and suppose that \(y\in\ker A(x)\cap\ker B(x)^*\).  Exactness of \eqref{eq:dual-resolution} over the point $[x]\in X$ represented by $x$ gives $y=B(x)c$ for some $c\in\C^\ell$. Hence,
\[
 0=B(x)^*B(x)c,
\]
and therefore
\[
 0=c^*B(x)^*B(x)c=\|B(x)c\|^2=\|y\|^2.
\]
Thus $y=0$. 
Proposition~\ref{prop:kernel-intersection} now gives the desired subspaces $D$ and $E$.
\end{proof}

We work on $X=\PP^3_{\C}$.  By the resolution criterion of Proposition~\ref{prop:resolution-criterion}, a witness pair $(D,E)$ for the two asymmetric $m=4$ families will follow once we produce a rank-two vector bundle $F$ on $X$, with the given parameters $(k,n,\ell)$, fitting into a linear resolution
\[
0\longrightarrow\OO(-2)^k
\longrightarrow\OO(-1)^n
\longrightarrow\OO^\ell
\longrightarrow F\longrightarrow0,
\qquad
H^0(F^\vee)=0.
\]

To build such an $F$, we first prove Proposition~\ref{prop:finite-resolution} which tells us that \emph{any} rank-two bundle on $X$ automatically has a resolution of this shape as soon as some of its cohomology groups vanish and its Euler characteristics in three low twists match the desired numbers. 
Next, we produce an actual bundle $F$ with the required properties, using the classical notion of \emph{natural cohomology} together with the existence theorem of Hartshorne and Hirschowitz (Theorem \ref{thm:natural-existence}). Matching Chern classes to $(k,n,\ell)$ reduces to a single integer parameter $\delta=\ell-k$, and the two asymmetric families of the Introduction turn out to be exactly the two residue classes of $\delta$ modulo $3$ for which the construction goes through. 

\subsection{From the $0$-regularity to a linear resolution}\label{subsec:finite-resolution}

Recall that a coherent sheaf $F$ on $X=\PP^{3}$ is \emph{$a$-regular} if $H^i(F(a-i))=0$ for $1\le i\le3$. 
We use this Castelnuovo--Mumford regularity through two standard facts, each proved in \cite[Tags 08A6 and 08A8]{Stacks}. 
\begin{enumerate}[label=(R\arabic*)]
\item \emph{Persistence}: if $F$ is $a$-regular then $F$ is $(a+1)$-regular.
\item \emph{Generation}: if $F$ is $a$-regular then $F(a)$ is globally generated, that is, the evaluation map $H^0(F(a))\otimes\OO_X\to F(a)$ is surjective.
\end{enumerate}
The cohomology of line bundles on $X=\PP^3$ is well-known  
(cf. \cite[Theorem III.5.1]{Hart} or \cite[Tag 01XS]{Stacks}): 
among $H^0(\OO_X(j)),\dots,H^3(\OO_X(j))$, only $H^0(\OO_X(j))$ for $j\ge0$ and $H^3(\OO_X(j))$ for $j\le-4$ can be nonzero.  In particular $\OO_X(-1),\OO_X(-2),\OO_X(-3)$ are acyclic.
We also use, without further comment, the long exact cohomology sequence attached to a short exact sequence of sheaves.

\begin{proposition}\label{prop:finite-resolution}
Let $F$ be a $0$-regular rank-two vector bundle on $X=\PP^3$.  
Let $k,n,\ell$ be nonnegative integers with $\ell-n+k=2$. 
If
\begin{equation}\label{eq:finite-euler}
 \chi(F)=\ell,\qquad
 \chi(F(1))=4\ell-n,\qquad
 \chi(F(2))=10\ell-4n+k,
\end{equation}
then there is an exact sequence
\begin{equation}\label{eq:F-resolution}
0\longrightarrow\OO(-2)^k
\longrightarrow\OO(-1)^n
\longrightarrow\OO^\ell
\longrightarrow F\longrightarrow0.
\end{equation}
\end{proposition}

\begin{proof}
Since $F$ is $0$-regular, $F$ is globally generated with $h^0(F)=\chi(F)=\ell$.  
Let
\begin{equation}\label{eq:K-eval}
0\longrightarrow K\longrightarrow\OO_X^\ell\longrightarrow F\longrightarrow0
\end{equation}
be the evaluation sequence.  
By construction, the induced map $H^0(\OO^\ell)\to H^0(F)$ is an isomorphism. Since $H^1(\OO_X)=0$, the long exact sequence forces $H^0(K)=H^1(K)=0$.  
By diagram chases after twisting by line bundles, we find that 
$H^1(K)=H^2(K(-1))=H^3(K(-2))=0$ and hence $K$ is $1$-regular.

By (R2), $K(1)$ is globally generated, and persistence (R1) together with the same long-exact-sequence argument, applied to \eqref{eq:K-eval} twisted by $1$, gives
\[
 h^0(K(1))=4\ell-h^0(F(1))=n.
\]
Its evaluation sequence, twisted back by $-1$, is
\begin{equation}\label{eq:L-eval}
0\longrightarrow L\longrightarrow\OO(-1)^n\longrightarrow K\longrightarrow0.
\end{equation}
By repeating the same argument, we find that $L$ is $2$-regular of rank
\[
 \rk L=n-(\ell-2)=k.
\]
Using $H^1(L(2))=H^1(K(2))=0$ from the regularity, we obtain
\[
 h^0(L(2))=4n-h^0(K(2))=4n-10\ell+h^0(F(2))=k.
\]
The globally generated bundle $L(2)$ thus has exactly its rank in global sections, so its evaluation map $\OO^k\to L(2)$ is a surjective map of bundles of equal rank, hence an isomorphism: locally it is a square matrix whose determinant is a nowhere-vanishing section of a line bundle, i.e.\ a unit.  This also covers the $k=0$ case trivially.  Consequently $L\simeq\OO(-2)^k$, and splicing \eqref{eq:K-eval} and \eqref{eq:L-eval} produces the resolution \eqref{eq:F-resolution}.
\end{proof}

\subsection{Stable bundles with natural cohomology and the parameter $\delta$}\label{subsec:natural-cohomology}
We first recall the following 
from \cite{HartshorneHirschowitz1982}. 
\begin{definition}
(1) A vector bundle $F$ on $X=\PP^3$ is said to have \emph{natural cohomology} if for every integer $u$, at most one of the cohomology groups $H^i(F(u))$ is nonzero.

(2) A rank-two vector bundle $F$ on $\PP^3$ is called \emph{stable} if $2\deg L>\deg \det(F)=c_1(F)$ for every surjection $F\to L$ onto a rank-one sheaf.
\end{definition}
It is obvious that if $F$ has natural cohomology, so does $F(a)$ for any integer $a$. Note that if $F$ is a stable rank-two bundle, so is $F(a)$ for any $a$ because $c_1(F(a))=c_1(F)+2a$ and any surjection $F(a)\to L$ gives a surjection $F\to L(-a)$.

\begin{lemma}\label{rem:stability-fact}
    If $F$ is a stable bundle on $X=\PP^3$ with $c_1(F)>0$, $H^0(F^\vee)=0$.
\end{lemma} 
\begin{proof}
    If $H^0(F^\vee)\ne 0$, there is a nonzero morphism  $F\to \OO_X$ whose image is a rank 1 sheaf $L$ with non-positive degree as it is a subsheaf of $\OO_X$. Then $0\ge 2\deg L>c_1(F)$ by stability. This contradicts our assumption $c_1(F)>0$.
\end{proof}

We use the following theorem of Hartshorne and Hirschowitz. 
\begin{theorem}\cite[Theorem~0.1]{HartshorneHirschowitz1982}\label{thm:natural-existence}
There exists a stable rank-two bundle $F$ on $\PP^3$ with natural cohomology and Chern classes $(c_1(F),c_2(F))=(\epsilon,d)$ whenever
\[
 \epsilon=0,\ d>0,
 \qquad\text{or}\qquad
 \epsilon=-1,\ d\ge6\text{ even}.
\]
\end{theorem}

For an integer $\delta\ge5$ with $3\nmid\delta$, put
\begin{equation}\label{eq:delta-parameters}
 n=\frac{\delta^2-4}{3},\qquad
 k=\frac{(\delta-1)(\delta-2)}6,\qquad
 \ell=\frac{(\delta+1)(\delta+2)}6.
\end{equation}
These are positive integers, with $\ell-k=\delta$ and $\ell-n+k=2$.  The Hilbert polynomial $\chi(F(u))$ of a rank-two bundle $F$ with 
Chern classes given by \eqref{eq:F-chern} is
\begin{equation}\label{eq:delta-hilbert}
 P_\delta(u)=\frac{(u+1)(u+\delta+1)(2u+\delta+2)}6.
\end{equation}
by Riemann--Roch. 
Direct substitution gives
\begin{align}
 P_\delta(-1)&=0,&P_\delta(-2)&=-k,\notag\\
 P_\delta(0)&=\ell,&P_\delta(1)&=4\ell-n,&
 P_\delta(2)&=10\ell-4n+k.\label{eq:delta-positive}
\end{align}

\begin{proposition}\label{prop:natural-resolution}
Let $\delta\ge5$, $3\nmid\delta$, and let $F$ be a stable rank-two bundle on $X$ with natural cohomology and
\begin{equation}\label{eq:F-chern}
 c_1(F)=\delta-2,\qquad
 c_2(F)=\frac{(\delta-1)(\delta-2)}3.
\end{equation}
Then $F$ is $0$-regular and satisfies \eqref{eq:finite-euler}. Therefore, we have the resolution \eqref{eq:F-resolution} with $H^0(F^\vee)=0$.
\end{proposition}

\begin{proof}
For every $j\ge0$,
$F(j)=F\otimes \OO(j)$ is stable with  
\[
 c_1\bigl(F(j)\bigr)=c_1(F)+2j=(\delta-2)+2j\ge \delta-2>0,
\]
using $\delta\ge5$. Hence $F(j)$ is stable with positive $c_1$, and Lemma~\ref{rem:stability-fact} gives
\begin{equation}\label{eq:dual-no-sections}
 H^0(F^\vee(-j))=0,\qquad \forall j\ge0.
\end{equation}
Next, natural cohomology together with $P_\delta(-1)=0$ forces $F(-1)$ to be acyclic: at most one of $H^0(F(-1)),\dots,H^3(F(-1))$ can be nonzero, and their alternating sum is $\chi(F(-1))=P_\delta(-1)=0$, so all four vanish. By the same token, if $H^2(F(-2))$ were nonzero, it would be the unique nonzero cohomology group of $F(-2)$, forcing $\chi(F(-2))\ge0$; but $\chi(F(-2))=P_\delta(-2)=-k<0$, a contradiction, so $H^2(F(-2))=0$.  Finally, Serre duality on $\PP^3$, 
gives
\[
 H^3(F(-3))\simeq 
 H^0(F^\vee(-1))^\vee=0,
\]
the last equality being the case $j=1$ of \eqref{eq:dual-no-sections}.  These three vanishings imply the $0$-regularity of $F$, while \eqref{eq:delta-positive} is exactly \eqref{eq:finite-euler}. Proposition~\ref{prop:finite-resolution} now gives the resolution \eqref{eq:F-resolution}, and the case $j=0$ of \eqref{eq:dual-no-sections} gives the vanishing $H^0(F^\vee)=0$.
\end{proof}

\begin{theorem}\label{thm:m4-delta}
For every integer $\delta\ge8$ with $3\nmid\delta$, the parameters \eqref{eq:delta-parameters} admit subspaces $D,E\subset\C^4\otimes\C^n$ of codimensions $k,\ell$ satisfying \eqref{eq:range-condition}.
\end{theorem}

\begin{proof}
We find a bundle $F'$ as in Theorem~\ref{thm:natural-existence} whose twist $F=F'(a)$ has exactly the Chern classes \eqref{eq:F-chern}.  Twisting a rank-two bundle by $\OO(a)$ changes $(c_1,c_2)=(\epsilon,d)$ to $(\epsilon+2a,\,d+a\epsilon+a^2)$, so we must solve $\epsilon+2a=\delta-2$ for one of the two values $\epsilon\in\{0,-1\}$ allowed by Theorem~\ref{thm:natural-existence}, with $a\in\mathbb Z$; this fixes $\epsilon$ by the parity of $\delta$, and then $a$ and the required $d=c_2(F')$:
\[
\begin{array}{c|ccc}
 &\epsilon=c_1(F')&a&d=c_2(F')\\ \hline
\delta\text{ even}&0&(\delta-2)/2&(\delta^2-4)/12\\
\delta\text{ odd}&-1&(\delta-1)/2&(\delta^2-1)/12
\end{array}
\]
the last column being the value of $d$ that $c_2(F)=d+a\epsilon+a^2=(\delta-1)(\delta-2)/3$ forces.

For even $\delta$, $a$ is an integer, and $\delta^2-4=(\delta-2)(\delta+2)$ is divisible by $4$ (both factors are even) and by $3$ (since $3\nmid\delta$ forces $\delta^2\equiv1\bmod3$, so $\delta^2-4\equiv0\bmod3$); thus $d$ is an integer, and $d\ge5$ once $\delta\ge8$.  This is the case $\epsilon=0$, $d>0$ of Theorem~\ref{thm:natural-existence}.

For odd $\delta$ coprime to $3$, $\delta^2-1=(\delta-1)(\delta+1)$ is a product of two consecutive even integers, hence divisible by $8$, and $3\nmid\delta$ again gives $\delta^2\equiv1\bmod3$, so $3\mid\delta^2-1$; thus $d$ is an even integer.  The case $\epsilon=-1$ of Theorem~\ref{thm:natural-existence} further requires $d\ge6$; since $\delta=9$ is excluded, the least admissible odd $\delta\ge8$ is $\delta=11$, giving $d=10\ge6$.

In both cases, Theorem~\ref{thm:natural-existence} supplies a stable bundle $F'$ with natural cohomology and Chern classes $(\epsilon,d)$.  Set $F=F'(a)$; twisting preserves both stability and natural cohomology, and by the twisting formulas above, $c_1(F)=\epsilon+2a=\delta-2$ and $c_2(F)=d+a\epsilon+a^2=(\delta-1)(\delta-2)/3$, exactly \eqref{eq:F-chern}.  Proposition~\ref{prop:natural-resolution} now supplies the resolution \eqref{eq:F-resolution} and $H^0(F^\vee)=0$, and Proposition~\ref{prop:resolution-criterion} converts these into the required subspaces $D,E$.
\end{proof}

\begin{proof}[Proof of Theorem~\ref{thm:main} for \eqref{eq:m4-family} with \(t\ge3\) and \eqref{eq:m4-second-family} with \(s\ge2\)]
For the first branch set $\delta=3t+1$, with $t\ge3$; for the second set $\delta=3s+2$, with $s\ge2$.
Substitution in \eqref{eq:delta-parameters} gives exactly \eqref{eq:m4-family} and \eqref{eq:m4-second-family}.  Conversely, every $\delta\ge8$ not divisible by $3$ has one of these two forms with the stated parameter bound.  
This proves Theorem~\ref{thm:main} for both families.
\end{proof}

The two Hilbert polynomials of \eqref{eq:delta-hilbert} are, on the two branches $\delta=3t+1$ and $\delta=3s+2$,
\begin{align*}
 p_t(u)&=P_{3t+1}(u)=\frac{(u+1)(u+3t+2)(2u+3t+3)}6,\\
 q_s(u)&=P_{3s+2}(u)=\frac{(u+1)(u+3s+3)(2u+3s+4)}6.
\end{align*}
The threshold $\delta\ge8$ in Theorem~\ref{thm:m4-delta} excludes exactly $\delta=5$ and $\delta=7$, the two finite cases taken up in Section~\ref{sec:4727-witness}: the conditional Proposition~\ref{prop:natural-resolution} still applies to these two numerical values, but the bundle existence it requires does not follow from Theorem~\ref{thm:natural-existence}, whose hypotheses fail at both.

\section{Explicit matrix constructions for the two remaining \(m=4\) cases}\label{sec:4727-witness}
In the previous section, we proved Theorem~\ref{thm:main} for \eqref{eq:m4-family} with \(t\ge3\) and \eqref{eq:m4-second-family} with \(s\ge2\).
The only remaining cases are \(t=2\) and \(s=1\), namely \((4,15,5,12)\) and \((4,7,2,7)\).
This section settles both by exhibiting the subspaces directly.

Throughout this section, we will use the following notation.
If $X=(X_{ij})_{1\le i,j\le4}$ is an $4n\times 4n$ matrix written as a $4\times4$ array of $n\times n$ blocks $X_{ij}$, define
\[
X^\Gamma=(X_{ji})_{1\le i,j\le4}.
\]
For a square matrix \(R\), we write \(R\succeq0\) if \(R\) is Hermitian positive semidefinite.
For Hermitian matrices \(P,Q\) of the same size, we write \(P\succeq Q\) if \(P-Q\succeq0\).

\begin{proposition}\label{prop:sos-criterion}
Let $n,k,\ell\ge1$ and set $N=4n$.  For
$A_i\in M_{k\times n}(\C)$ and
$B_i\in M_{n\times\ell}(\C)$, $1\le i\le4$, write
\(A(x)=\sum_{i=1}^4x_iA_i\) and \(B(x)=\sum_{i=1}^4x_iB_i\), and define $\bfA\in M_{k\times N}(\C)$ and $\bfB\in M_{N\times\ell}(\C)$ as in \eqref{eq:block-matrices}.

Suppose that $\rk \bfA=k$ and $\rk \bfB=\ell$, and that there exist
$\delta>0$ and $R\succeq0$ such that
\begin{equation}\label{eq:sos-condition}
 (\bfA^*\bfA)^\Gamma
 +\bfB\bfB^*
 -R^\Gamma
 \succeq
 \delta I_N.
\end{equation}
Then the matrices \(A_i,B_i\) determine subspaces \(D,E\) of codimensions \(k,\ell\), respectively, satisfying the range condition \eqref{eq:range-condition}.
\end{proposition}

\begin{proof}
By \eqref{eq:block-rank-conditions}, the rank assumptions imply that the matrices \(A_i,B_i\) determine subspaces \(D,E\) of codimensions \(k,\ell\), respectively.
It remains to verify the range condition \eqref{eq:range-condition}.
Set
\[
T(x)=[\,x_1I_n\mid x_2I_n\mid x_3I_n\mid x_4I_n\,]\in M_{n\times N}(\C).
\]
Then \(T(x)\bfB=B(x)\), and hence \(T(x)\bfB\bfB^*T(x)^*=B(x)B(x)^*\).

Next, the $(i,j)$-block of $\bfA^*\bfA$ is $A_i^*A_j$.
Therefore, the $(i,j)$-block of
$(\bfA^*\bfA)^\Gamma$ is $A_j^*A_i$, and
\[
T(x)(\bfA^*\bfA)^\Gamma T(x)^*
=
\sum_{i,j=1}^4 \overline{x_i}x_j\,A_i^*A_j
=
A(x)^*A(x).
\]

More generally, if $X=(X_{ij})$ is written in $n\times n$ blocks, then
\[
T(x)X^\Gamma T(x)^*
=\sum_{i,j=1}^4 \overline{x_i}x_j\,X_{ij}
=T(\overline{x})XT(\overline{x})^*.
\]
Also,
\[
T(x)T(x)^*
=
\sum_{i=1}^4 |x_i|^2I_n
=
\lVert x\rVert^2I_n.
\]

The transformation $X\mapsto T(x)XT(x)^*$ preserves Hermitian positive semidefiniteness, since
\[
v^*T(x)XT(x)^*v=(T(x)^*v)^*X(T(x)^*v)\ge0
\]
whenever $X\succeq0$.
Applying it to \eqref{eq:sos-condition} therefore gives
\[
A(x)^*A(x)+B(x)B(x)^*
-T(\overline{x})RT(\overline{x})^*
\succeq
\delta\lVert x\rVert^2I_n.
\]
Since $R\succeq0$, we also have
$T(\overline{x})RT(\overline{x})^*\succeq0$.
Hence
\begin{equation}\label{eq:target-positivity}
A(x)^*A(x)+B(x)B(x)^*
\succeq
\delta\lVert x\rVert^2I_n
\qquad\text{for every }x\in\C^4.
\end{equation}

If \(x\ne0\) and \(y\in\ker A(x)\cap\ker B(x)^*\), then applying \eqref{eq:target-positivity} to \(y\) gives \(0\ge\delta\lVert x\rVert^2\lVert y\rVert^2\), and hence \(y=0\).
Thus, \(\ker A(x)\cap\ker B(x)^*=0\) for every \(x\ne0\), and Proposition~\ref{prop:kernel-intersection} applies.
\end{proof}

Note that \eqref{eq:target-positivity} is a matrix inequality that must hold for every $x\in\C^4$. In contrast, \eqref{eq:sos-condition} replaces it by a single positive semidefiniteness condition on a constant $4n\times4n$ matrix.
This is useful because the latter can be checked by a straightforward matrix calculation.

We will choose $A(x)$ in a simple structured form. Given $n$ and $k$, let
$A(x)$ be the $k\times n$ band matrix
\[
\bigl(A(x)\bigr)_{a,c}
= \sum_{i=1}^{4} x_i \delta_{c,a+d_i},
\qquad
(d_1,d_2,d_3,d_4)=(0,1,k+1,n-k),
\]
where $1\le a\le k$ and $1\le c\le n$.
For $(n,k)=(7,2)$, $(d_1,d_2,d_3,d_4) = (0,1,3,5)$ and

\begin{equation}\label{eq:A-4727}
 A(x)=
 \begin{pmatrix}
 x_1&x_2&0&x_3&0&x_4&0\\
 0&x_1&x_2&0&x_3&0&x_4
 \end{pmatrix}.
\end{equation}
Its $2\times2$ minors include $x_1^2,x_2^2,x_3^2,x_4^2$, so $A(x)$ has
rank two for every $x\ne0$.

For $(n,k)=(15,5)$, $(d_1,d_2,d_3,d_4)=(0,1,6,10)$, and
{\scriptsize
\setlength{\arraycolsep}{3pt}
\begin{equation}\label{eq:A-415512}
A(x)=
\begin{pmatrix}
x_1&x_2&0&0&0&0&x_3&0&0&0&x_4&0&0&0&0\\
0&x_1&x_2&0&0&0&0&x_3&0&0&0&x_4&0&0&0\\
0&0&x_1&x_2&0&0&0&0&x_3&0&0&0&x_4&0&0\\
0&0&0&x_1&x_2&0&0&0&0&x_3&0&0&0&x_4&0\\
0&0&0&0&x_1&x_2&0&0&0&0&x_3&0&0&0&x_4
\end{pmatrix}.
\end{equation}
}
The $5\times5$ minors on columns $1,\dots,5$, $2,\dots,6$,
$7,\dots,11$, and $11,\dots,15$ are respectively
$x_1^5,x_2^5,x_3^5,x_4^5$. 
Hence $\rk A(x)=5$ for every $x\ne0$.

The constructions above therefore fix \(\bfA\) in both cases.
To complete the proof of Theorem~\ref{thm:main}, it remains to exhibit \(\bfB\) and \(R\succeq0\) satisfying the hypotheses of Proposition~\ref{prop:sos-criterion} in the two remaining cases.

\begin{proof}[Proof of Theorem~\ref{thm:main} for \((4,7,2,7)\) and \((4,15,5,12)\)]
The coefficient matrices $A_1,\ldots,A_4$ are given explicitly in \eqref{eq:A-4727} and \eqref{eq:A-415512}, respectively.
Since the block matrix \(\bfB\) and the factor \(L\) are large and dense, we record their exact entries in the verification scripts available at \url{https://github.com/chlee-0/krawtchouk-range-criterion}.

For $(m,n,k,\ell)=(4,7,2,7)$, explicit matrices $B_1,\ldots,B_4\in M_7(\mathbb Z[i])$ and $L\in M_{28\times9}(\mathbb Q(i))$ are given in \href{https://github.com/chlee-0/krawtchouk-range-criterion/blob/main/verify_4727.py}{\texttt{verify\_4727.py}}.
Setting $R=LL^*\succeq0$ and $\delta=1/5000$, the script verifies, using exact arithmetic, that $\rk\bfA=2$, $\rk\bfB=7$, and that the matrix inequality \eqref{eq:sos-condition} holds.

For $(m,n,k,\ell)=(4,15,5,12)$, explicit matrices $B_1,\ldots,B_4\in M_{15\times12}(\mathbb Z[i])$ and $L\in M_{60\times25}(\mathbb Q(i))$ are given in \href{https://github.com/chlee-0/krawtchouk-range-criterion/blob/main/verify_415512.py}
{\texttt{verify\_415512.py}}.
Setting $R=LL^*\succeq0$ and $\delta=1/20000$, the script verifies, again using exact arithmetic, that $\rk\bfA=5$, $\rk\bfB=12$, and that \eqref{eq:sos-condition}
holds.

Thus all the hypotheses of Proposition~\ref{prop:sos-criterion} are satisfied in both cases, and the two assertions follow.
\end{proof}

\begin{remark}
We briefly describe how the matrices used above were found.
With \(\bfA\) fixed, we searched numerically for \(\bfB\) together with a factor \(L\), where \(R=LL^*\).
For a candidate pair \((\bfB,L)\), we measured
\[
\mu(\bfB,L)=\lambda_{\min}\!\left((\bfA^*\bfA)^\Gamma+\bfB\bfB^*-(LL^*)^\Gamma\right).
\]
Thus, \(\mu(\bfB,L)>0\) is precisely the desired positivity condition, with any \(0<\delta\le \mu(\bfB,L)\).
A direct search for a pair with \(\rk\bfB=\ell\) and \(\mu(\bfB,L)>0\) was ineffective.
We therefore began with a matrix \(\widetilde{\bfB}\) having one additional column.
Starting from a pair with \(\mu(\widetilde{\bfB},L)>0\), we gradually forced the smallest singular value of \(\widetilde{\bfB}\) toward zero, repeatedly reoptimizing \(\widetilde{\bfB}\) and \(L\) while keeping \(\mu>0\).
When the smallest singular value approaches zero, the extra column becomes redundant and, at successful endpoints, the resulting matrix has rank $\ell$ while still satisfying $\mu>0$.
The resulting numerical certificate was then simplified to Gaussian-integer and Gaussian-rational data and verified exactly.
\end{remark}

\section*{Declaration of generative AI use}

During the preparation of this manuscript, the authors used OpenAI ChatGPT to explore mathematical ideas, to develop and debug computational code, and to assist with editorial and formatting checks. The authors critically reviewed the AI-generated suggestions and computational results.

\section*{Acknowledgements}

C.-h. Lee was supported by a KIAS Individual Grant (AP067301) via the Center for Artificial Intelligence and Natural Sciences at Korea Institute for Advanced Study.


\begin{thebibliography}{99}

\bibitem{Stacks}
The Stacks Project Authors,
\emph{The Stacks Project}, Tags 01XS, 08A6, and 08A8,
\url{https://stacks.math.columbia.edu}.

\bibitem{Hart}
R.~Hartshorne, \emph{Algebraic geometry}, Graduate Texts in Mathematics 52, Springer 1977. 

\bibitem{HartshorneHirschowitz1982}
R.~Hartshorne and A.~Hirschowitz,
\emph{Cohomology of a general instanton bundle},
Ann. Sci. \'{E}cole Norm. Sup. (4) \textbf{15} (1982), no.~2, 365--390.
\href{https://doi.org/10.24033/asens.1429}{doi:10.24033/asens.1429}.

\bibitem{HeoKiem2019}
J.~Heo and Y.-H.~Kiem,
\emph{On characterizing integral zeros of Krawtchouk polynomials by quantum entanglement},
Linear Algebra Appl. \textbf{567} (2019), 167--179.
\href{https://doi.org/10.1016/j.laa.2019.01.005}{doi:10.1016/j.laa.2019.01.005}.

\bibitem{Horodecki1997}
P.~Horodecki,
\emph{Separability criterion and inseparable mixed states with positive partial transposition},
Phys. Lett. A, \textbf{232} (1997), 333--339.

\bibitem{KiemKyeLee2011}
Y.-H.~Kiem, S.-H.~Kye, and J.-S.~Lee,
\emph{Existence of product vectors and their partial conjugates in a pair of spaces},
J. Math. Phys. \textbf{52} (2011), 122201.

\bibitem{Kra92} 
I. Krasikov and Y. Roditty.
\emph{Switching reconstruction and diophantine equations.}
J. Comb. Theory, Ser. B, \textbf{54} (1992), 189--195.

\bibitem{Kra96} 
I.~Krasikov and S.~Litsyn.
\emph{On integral zeros of Krawtchouk polynomials},
J. Comb. Theory, Ser. A, \textbf{74} (1996), 71--99.

\bibitem{LKHC2001}
M.~Lewenstein, B.~Kraus, P.~Horodecki, and J.I.~Cirac,
\emph{Characterization of separable states and entanglement witnesses},
Phys. Rev. A, \textbf{63} (2001), 044304.

\bibitem{Peres1996}
A.~Peres,
\emph{Separability criterion for density matrices},
Phys. Rev. Lett., \textbf{77} (1996), 1413--1415.

\bibitem{Vakil}
R.~Vakil, \emph{The rising sea: foundations of algebraic geometry}, Princeton University Press, 2025. 

\end{thebibliography}
\end{document}